\documentstyle[aas2pp4]{article}

\makeatletter
\renewenvironment{deluxetable}[1]{\def\pt@format{\string#1}%
\set@tblnotetext\global\pt@ncol=0\global\pt@column=0\global\pt@page=1%
\def\pt@addcol{\global\advance\pt@ncol by\@ne}}%
{\pt@width\wd\pt@box\box\pt@box\spew@ptblnotes%
\typeout{Page \the\pt@page\space of table \thetable\space has been set to
width \the\pt@width\space with \the\pt@nlines\space lines per page}%
\endcenter\end@float}
\def\startdata{\pt@line=0\pt@calcnlines%
\ifdim\pt@width>\z@\def\@halignto{to \pt@width}\else\def\@halignto{}\fi%
\let\fnum@table=\fnum@ptable\set@mkcaption%
\@float{table}\center\caption{\pt@caption}\leavevmode%
\setbox\pt@box=\pt@tabular{\pt@format}\pt@head}
\def\thebibliography{\subsection*{REFERENCES}
\list{}{\labelwidth3em\leftmargin\labelwidth\labelsep\z@\parsep\z@
\itemsep\z@\itemindent-3em\usecounter{enumi}}
\def\refpar{\relax}
\def\newblock{\hskip .11em plus .33em minus .07em}
\sloppy\clubpenalty4000\widowpenalty4000
\sfcode`\.=1000\relax}
\makeatother

\def\rxjw{RX~J185635$-$3754}
\def\rxj{RX~J0720.4$-$3125}
\def\Xray{\hbox{X-ray}}

\begin{document}

\title{Optical Observations of the Isolated Neutron Star \rxj}
\author{S. R. Kulkarni}
\affil{Palomar Observatory 105-24, California Institute of Technology, 
Pasadena, CA 91125, USA}
\authoremail{srk@astro.caltech.edu}
\and
\author{M. H. van Kerkwijk}
\affil{Institute of Astronomy, Madingley Road, Cambridge CB3 0HA, UK}
\authoremail{mhvk@ast.cam.ac.uk}

\begin{abstract}
\rxj\ is an unidentified bright soft \Xray\ source which shows
pulsations at a 8.39\,s period and has a thermal spectrum.  We present
deep B and R band images of its \Xray\ localization.  We find one
possible counterpart in the \Xray\ error box, with magnitudes
$B=26.6\pm0.2$ and $R=26.9\pm0.3$.  The very high X-ray to optical
flux ratio confirms that this object is an isolated neutron star.  We
discuss possible models and conclude that only two are consistent with
the data and at the same time are able to draw from a large enough
population to make finding one nearby likely.  In our opinion the
second criterion provides a stringent constraint but appears to have
been ignored so far.  The first model, suggested earlier, is that
\rxj\ is a  weakly magnetized neutron star accreting from the
interstellar medium.  The second is that it is a relatively young,
highly magnetized neutron star, a ``magnetar'', which is kept hot by
magnetic field decay.
\end{abstract}


\section{Introduction\label{sec:intro}}

The population of defunct radio pulsars far exceeds that of active
ones.  It is believed that the Galaxy has about $2\,10^5$ radio
pulsars.  The neutron-star birth rate is estimated to be between one
per 30 yr to one per 100 yr.  Assuming a constant pulsar production
rate and an age of the disk of $10^{10}$\,yr, one infers a Galactic
neutron star population of $\sim\!2\,10^8$ -- three orders of
magnitude larger than that of the active radio pulsar population.

It is not easy to detect old neutron stars.  While the nearest
few intermediate-age pulsars can be identified by their cooling
radiation, which peaks in the soft \Xray/EUV band, the defunct pulsars
will have become too cool to be observable.  A small fraction of them,
however, may be in a position to accrete matter from the interstellar
medium.  These will then get reheated and reappear in the \Xray\ sky.

Quite independent of this discussion there has been a growing
recognition of a population of highly magnetized neutron stars.  The
circumstantial evidence for this class comes from studies of soft
gamma-ray repeaters (SGRs) and long-period pulsars in supernova
remnants (Vasisht \& Gotthelf \cite{vasig:97}).  Thompson \& Duncan
(\cite{thomd:95}) have introduced the term ``magnetars'' for neutron
stars with field strengths significantly larger than $10^{12}$\,G, the
typical field strength inferred for radio and \Xray\ pulsars.  The
birthrate of SGRs has been estimated to be roughly 10\% that of the
ordinary pulsars (Kulkarni \& Frail \cite{kulkf:93}; Kouveliotou et
al.\ \cite{kouv&a:94}). 
 
The relevance of magnetars to the discussion at hand is as follows.
Unlike the situation for ordinary neutron stars, magnetic field decay
is expected to be significant in highly magnetized neutron stars.
This decay could reheat the magnetar (Thompson \& Duncan
\cite{thomd:96}), making it hotter than an ordinary neutron star, and
thus brighter in soft X rays. 

Two of the best candidates for this general class of neutron stars have
emerged from the ROSAT mission: \rxjw\ (Walter, Wolk, \& Neuh\"auser
\cite{waltwn:96}) and \rxj\ (Haberl et al.\ \cite{habe&a:97}). Both are
bright ROSAT objects with very soft \Xray\ spectra.  Walter \& Matthews
(\cite{waltm:97}) have provided compelling evidence for the
identification of a faint blue optical counterpart of \rxjw.  In this
{\em Letter}, we present deep B and R observations of the localization
of \rxj.

\section{RX~J0720.4$-$3125\label{sec:rxj}}

The basic \Xray\ properties are described by Haberl et al.\
(\cite{habe&a:97}).  A brief summary now follows.  With a PSPC count
rate of 1.67\,s$^{-1}$ the source is among the brighter in the ROSAT
All-Sky Survey.  The PSPC spectrum is fitted very well by a black body
with $kT_{\rm{}eff}=79\pm4$\,eV absorbed by a column density,
$N_H=(1.3\pm0.3)\,10^{20}{\rm\,cm}^{-2}$.  The observed flux in the
0.1--2.4 keV ROSAT band is
$1.15^{+0.3}_{-0.14}\,10^{-11}{\rm\,erg\,cm^{-2}\,s^{-1}}$. Using
PIMMS (Portable, Interactive Multi-Mission Simulator;
http://legacy.gsfc.nasa.gov), we estimate, for the black-body model
mentioned above, an unabsorbed flux of
$1.8\,10^{-11}{\rm\,erg\,cm^{-2}\,s^{-1}}$ in the range 0.010--2.4
keV.  

Perhaps the most remarkable fact about the source is neither its high
count rate nor its soft \Xray\ spectrum, but the presence of
pulsations with a 8.39\,s period.  The period is very stable: from
observations done over a three year period Haberl et al.\ find a
period derivative $\dot{P}=(-2.6\pm3.4)\,10^{-12}{\rm\,s\,s^{-1}}$.
This argues for the period to be the rotation period of the star.  The
shortness of the period rules out all models save those involving a
neutron star or a white dwarf.

Haberl et al.\ found no optical counterpart within the ROSAT HRI error
circle and noted a \Xray\ to optical flux of $\gtrsim\!500$.  Such a
high value is barely compatible with a model involving an accreting low
mass \Xray\ binary and certainly excludes any model based on an
accreting white dwarf binary.  We note that an isolated hot white dwarf
(planetary nebula nucleus) would be incompatible with the high inferred
black-body temperature.

\section{Optical Observations\label{sec:obs}}

We imaged the field containing \rxj\ with the Low Resolution Imaging
Spectrograph (Oke et al.\ \cite{oke&a:95}), mounted at the Cassegrain
focus of the Keck~II telescope, on the nights of November 28 and 29,
1997 (UT).  The detector was a $2048\times2048$ pixel Loral CCD which
was read out with two amplifiers.  The skies appeared to be clear on
the second night but the first night was plagued by cirrus.  A log of
the observations is given in Table~\ref{tab:obs-log}.

\begin{deluxetable}{lcrrcc}
\tablewidth{\hsize}
\tablecaption{Log of Observations\label{tab:obs-log}}
\tablehead{
\colhead{Field}&
\colhead{Filter}&
\colhead{UT}& 
\colhead{$t_{\rm exp}$}& 
\colhead{$\sec z$}}
\startdata
RX J0720.4$-$3125&   R&  12:44& $2\times300$& 1.60\nl
&                    B&  12:58& $4\times600$& 1.60\nl
&                    R&  13:47& $3\times300$& 1.63\nl
\tablevspace{2mm}
RX J0720.4$-$3125&   B&  12:27& $3\times900$& 1.60\nl
&                    R&  13:18& $7\times300$& 1.62\nl
PG 2336+004&         B&  08:28& 20&           1.50\nl
{\em (A,B)}&         R&  08:32& 5&            1.52\nl
SA 95&               R&  08:38& 5&            1.10\nl
{\em (41,42)}&       B&  08:45& 20&           1.10\nl
PG 1047+003&         R&  14:08& 5&            1.33\nl
{\em (PG,A,B,C)}&    B&  14:10& 20&           1.33\nl
Ruben 149&           B&  15:43& 10&           1.35\nl
{\em (Ru,B,C,E,F,G)}&R&  15:45& $2\times2$&   1.36\nl
\enddata
\tablecomments{The UT date is 1997 Nov.\ 28 for the first three
entries, 1997 Nov.\ 29 for all others.  All exposure times are in
seconds.  The standard fields are from Landolt (\cite{land:92}); below
the name the stars actually used for the photometric calibration are
listed.  The seeing (as measured in the R-band) on the first night was
1\arcsec\ and improved to 0\farcs8 on the second night.}
\end{deluxetable}

During the second night, four flux-standard fields from Landolt
(\cite{land:92}) were observed (defocusing the telescope to avoid
saturation of pixels).  We took care to place our targets and most of
the standards on the same side of the CCD chip, thereby ensuring that
the same amplifier was used.  From the standard stars, which had a
range in $B-R$ from $-0.4$ to 1.7, we determined the color and
extinction coefficients.  The root-mean-square residual around the
calibration is about 0.02\,mag in both B and~R. 
%

\begin{figure*}
{\centering\leavevmode\epsfxsize0.87\hsize\epsfbox{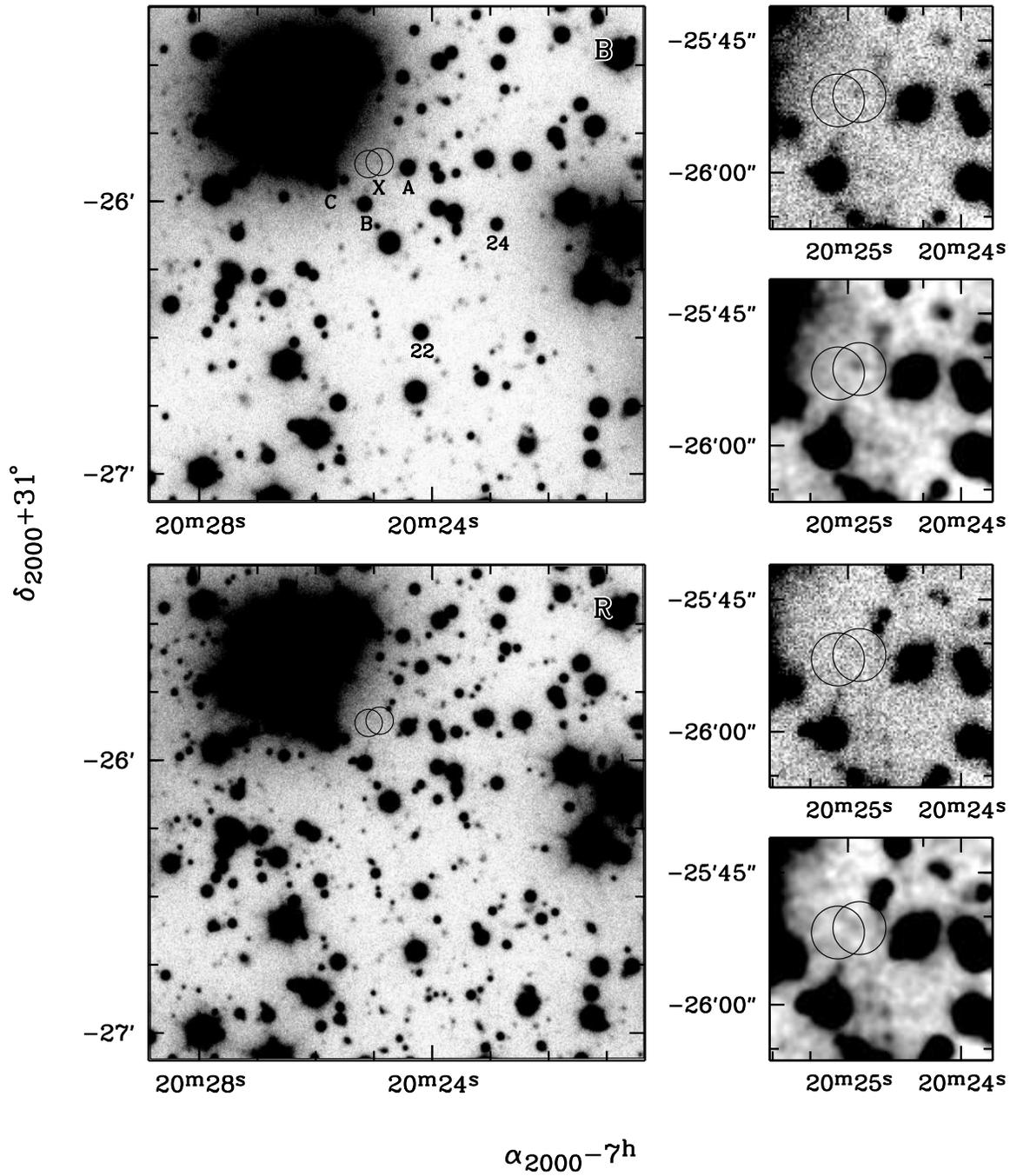}}
\caption[]{Summed B (top) and R (bottom) images of the localization of
\rxj.  The left-hand panels show an overview.  The two circles
represent the two HRI \Xray\ positions derived by Haberl et al.\
(1997).  Stars mentioned in the text are labeled below their image in
the top panel.  For each band, the region around the localization is
shown enlarged on the right-hand side.  For these enlargements, the
local sky was subtracted (determined by fitting a two-dimensional,
second-order polynomial to the image with stars masked out).  In the
lower panels, the image has furthermore been convolved with a Gaussian
with a width approximately matching the seeing (FWHM of four pixels --
0\farcs85).\label{fig:images}}
\end{figure*}

The reduction was done using MIDAS.  Individual frames were
bias-subtracted and then flat-fielded using twilight flats.  For the
\rxj\ images, a fixed region around the target's position was
extracted and cleaned of cosmic rays, and these images were added
together by band and by night.  The result is shown in
Figure~\ref{fig:images} (Plate).  Overdrawn are the two \Xray\
position estimates from Haberl et al.\ (\cite{habe&a:97}), using
astrometry relative to the USNO-A1.0 catalogue (Monet et al.\
\cite{mone&a:96}); see also Table~\ref{tab:phot-astr}.
Note that these are close to, but not exactly at,
the position shown in Fig.~7 of Haberl et al.  We do not know the
reason for this discrepancy.  

\begin{deluxetable}{lllcc}
\tablewidth{\hsize}
\tablecaption{Photometry and Astrometry of the Candidate Counterpart
of \rxj\ and Selected Stars in the Field\label{tab:phot-astr}}
\tablehead{
\colhead{Star}&
\colhead{B}&
\colhead{R}& 
\colhead{$\alpha_{\rm J2000}$}& 
\colhead{$\delta_{\rm J2000}$}}
\startdata
 X&   26.6(2)& 26.9(3)& 7 20 24.92&  $-$31 25 50.9\nl
 A&   20.52&   19.42&   7 20 24.40&  $-$31 25 52.3\nl
 B&   20.79&   19.62&   7 20 25.15&  $-$31 26 00.3\nl
 C&   21.23&   18.85&   7 20 25.76&  $-$31 25 55.6\nl
22&   20.37&   19.22&   7 20 24.18&  $-$31 26 28.4\nl
24&   21.49&   20.13&   7 20 22.88&  $-$31 26 04.6\nl
\enddata
\tablecomments{The uncertainty in the photometry for star X is
indicated by the number in brackets.  For the other stars, it is
$\lesssim\!0.02\,$mag in both B and R.  Comparing our measured $B-R$
to the main sequence colors (Johnson \cite{john:66}) we conclude that
the spectral types of A and B are both $\sim$\,G2 -- quite consistent
with what is indicated by the spectra of Haberl et
al.\ (\cite{habe&a:97}).  Haberl et al.\ classified star~C as M based
on its spectrum -- from both color and spectrum, we believe it is a
little earlier, $\sim$\,K7.  The V magnitudes inferred using these
spectral types are consistent with the range quoted by Haberl et al.
The uncertainty in the astrometry is dominated by the systematic
uncertainty in the USNO-A1.0 catalogue.  }
\end{deluxetable}

As can be seen in the figure, only one source (hereafter, ``X'') is
found within the union (and intersection) of the two HRI error
circles.  It is well detected in the B image but barely detected in
the R band.  
Scattered light from some nearby stars result in a strong gradient
in the local sky. This made
application of the standard photometric packages difficult.  We
proceeded by using a simplified point-spread function fitting, as
follows.  First, we used aperture photometry to measure the magnitudes
for four relatively isolated, brighter ``secondary'' stars (22, 24, A,
B; see Figure~\ref{fig:images} and Table~\ref{tab:phot-astr}).  Next,
we extracted the stellar images of these four stars, as well as those
of the candidate X and of star C, within a $21\times21$ pixel
($4\farcs5\times4\farcs5$) region centered on each source, and fitted
these to a 2-dimensional Gaussian on top of a plane with an arbitrary
tilt.  From the fit, we determined the average FWHM for the four
secondary stars.  We then refitted all objects keeping the FWHM fixed
at the average, and used the amplitudes of the Gaussians to determine
relative magnitudes.  
Finally, the magnitudes of stars X
and C were measured relative to the secondary stars which in turn were
calibrated using the solution found from the secondary stars. 

In summary, our observations have identified only one source, star X,
within the HRI localization region.  This source is a plausible
optical counterpart of \rxj.  Its magnitudes are $B=26.6\pm0.2$ and
$R=26.9\pm0.3$ (Table~\ref{tab:phot-astr}).  It is moderately blue,
$B-R=-0.3\pm0.4$.

\section{The Nature of \rxj\label{sec:nature}}

A straightforward conclusion we can draw from our observations is that
the ratio of the \Xray\ to the optical flux\footnote{In computing the
\Xray\ to optical flux it is traditional to compute the optical flux
using the following relation:
$f_V=10^{-0.4(V+13.42)}{\rm\,erg\,cm^{-2}\,s^{-1}}$; see Maccacaro et
al.\ (\cite{macc&a:88}).  We assumed $V\simeq B$ for source X.},
$f_X/f_V\gtrsim 2\,10^5$. The approximate equality is applicable if X
is the optical counterpart of \rxj.  The high value for the ratio
rules out the possibility that \rxj\ is an accreting neutron star
binary, leaving only a single neutron star as a viable option.  The
\Xray\ spectrum is highly suggestive of thermal emission from a hot
surface.  If so, $f_X$ constrains the distance to the source to be
less than $450R_6$\,pc; here $R_6$ is the radius of the neutron star
in units of $10^6$ cm.



\rxj\ must be nearby since the inferred column density is rather
small, $1.3\,10^{20}$ cm$^{-2}$. As pointed out by Haberl et al.\
(\cite{habe&a:97}), the source is in the general direction of the open
cluster Collinder~140 (Claria \& Rosenzweig \cite{clarr:78}).
FitzGerald, Harris \& Miller (\cite{fitzhm:80}) carried out extensive
spectroscopic observations and derive $E_{B-V}=0.04$, a distance of
$410\pm30$\,pc, and an age of $20\pm6$\,Myr.  Using the usual relation
between dust and gas, the inferred total column density is
$2.8\,10^{20}{\rm\,cm^{-2}}$. \rxj\ is within the confines of the
cluster , and thus, following Haberl et al.\ (\cite{habe&a:97}), we
place an upper limit of 410\,pc on the distance to \rxj.

Accepting the conclusion that \rxj\ is an isolated and nearby neutron
star, we now turn to the question of why its surface is hot.

\subsection{A Young Neutron Star}


In this model, \rxj\ is a nearby young neutron star.  According to
Umeda et al.\ (\cite{umed&a:93}), an isolated neutron star even with
internal frictional heating will cool down to $kT=80$\,eV at an age
$\tau_{\rm{}cool}=4\,10^5$ yr.  For a given neutron star birthrate
$B_{\rm{}ns}$, the expected number of neutron stars with an age less
than $\tau_{\rm{}cool}$ and within a distance projected on the
Galactic plane less than $\rho_{\rm{}lim}$ is
$B_{\rm{}ns}\tau_{\rm{}cool}\pi\rho_{\rm{}lim}^2$.  We make the
assumption that most neutron stars are born as ordinary pulsars. Lyne
et al.\ (\cite{lyne&a:98}) estimate the local birth rate of pulsars
with a 400-MHz luminosity above 1\,mJy\,kpc$^2$ as
$2.8\pm1.7{\rm\,Myr^{-1}\,kpc^{-2}}$.  Taking into account that many
pulsars will not be beamed towards us, they derive a total birth rate
of $10\pm6{\rm\,Myr^{-1}\,kpc^{-2}}$.  Lyne et al.\
claim that the luminosity function flattens below  
20\,mJy\,kpc$^2$. If so, $B_{\rm ns}$ is well constrained.
Hence, for
$\tau_{\rm{}cool}=0.4$\,Myr, $\rho_{\rm{}lim}=0.4$\,kpc, we expect a
total of 1.2--3.2 young nearby pulsars showing strong thermal
emission.  Geminga, at 160\,pc, is perhaps the nearest example of a
middle-aged cooling neutron star (Caraveo et al.\ \cite{cara&a:96};
Halpern \& Wang \cite{halpw:97}), while PSR~B0656+14 (Finley,
\"Ogelman, \& Kizilo\v{g}lu \cite{finlok:92}) and PSR~B1055-52
(\"Ogelman \& Finley \cite{ogelf:93}), both at $\sim\!500\,$pc, are
examples of neutron stars whose radio emission is beamed towards us.

A problem for the young-pulsar model, however, is the very long spin
period of 8.39\,s.  There is little evidence that normal pulsars are
born with such long periods, and it would have taken too long to spin
down from an initially short period. Thus, we consider this model to
be unlikely.

\subsection{A Neutron Star Accreting from the ISM}

In this picture, already extensively discussed in the literature
(Konenkov \& Popov \cite{konep:97}; Wang \cite{wang:97}; Haberl et
al.\ \cite{habe&a:97}), it is assumed that the neutron star is heated
by matter being accreted from the ISM.  An advantage compared to the
young pulsar model is that one can draw upon the entire old neutron
star population. The principal uncertainty is whether all the
conditions for accretion are met. These include: (1) the unknown and
small filling factor of moderately dense interstellar medium; (2) the
small fraction of neutron stars with sufficiently low velocity
($\lesssim\!10\%$; Cordes \& Chernoff \cite{cordc:97}); and (3) the
combination of dipole field strength and spin period needed to accrete
matter.  Accretion can proceed only if the Alfv\'en radius (the radius
at which accretion flow pressure balances the magnetospheric pressure)
is inside the corotation radius.  For \rxj, Wang (\cite{wang:97})
infers $B<10^{10}$\,G.  Thus, if the field originally was similar to
that observed in present-day young neutron stars, it must have decayed
substantially.  There is little evidence for any decay in ordinary
pulsars.  Indeed, Kulkarni \& Anderson (\cite{kulka:96}) argue that
the existence of long period pulsars in clusters is incompatible with
a field decay by more than a factor of 10 over a Hubble time.  For
this reason we neglect this model.

Haberl et al.\ (\cite{habe&a:97}) suggested an interesting possibility
that might alleviate points (2) and (3) mentioned above, namely that
\rxj\ is an evolved member of the group of so-called anomalous \Xray\
pulsars (Mereghetti \& Stella \cite{meres:95}).  These systems have
similar pulse periods, and are found close to the plane, often
associated with young SNRs.  If these systems are formed in
common-envelope evolution of high-mass \Xray\ binaries, as suggested
by van Paradijs et al.\ (\cite{vpartvdh:95}), their space velocities
are likely low, and their magnetic fields may be reduced due to
accretion.  For a birth rate of anomalous pulsars of about one per
$3\,10^3$\,yr (van Paradijs et al.\ \cite{vpartvdh:95}), there should
be $\sim\!1.6\,10^3$ descendants within $\rho_{\rm{}lim}<0.4$\,kpc.
If all of these have suitably low velocity, the fraction visible as
X-ray sources will depend only on the filling factor $f$ of dense
($\gtrsim\!1{\rm\,cm}^{-3}$) interstellar medium. Even for an
unrealistic $f\sim0.01$ we expect to see many accreting systems.
Thus, this is a plausible model, subject only to the assumed birthrate
of anomalous pulsars and whether they indeed evolve to low velocity,
low magnetic field, single neutron stars.

\subsection{A Magnetar\label{sec:magnetar}}

The long period of 8.39\,s is expected if the object is not a normal
young pulsar, but a magnetar. This possibility is briefly mentioned by
Wang (\cite{wang:97}) and considered in more detail by Heyl \&\
Hernquist (\cite{heyl:98}).  In the framework of the rotating magnet
(vacuum) model for pulsars, one has $P\dot{P}\propto{}B_{\rm{}d}^2$;
here $P$ is the rotation period and $B_{\rm{}d}$ the strength of the
dipole component of the magnetic field.  Assuming that the neutron
star was born with a period much shorter than the current period we
find $\dot{P}\equiv P/\tau=3\,10^{-13}\tau_6^{-1}{\rm\,s\,s^{-1}}$ where
the age of the magnetar is $\tau=10^6\tau_6$\,yr.  This leads to
$B_d\sim5\tau_6^{-1/2}\,10^{13}$\,G.

Earlier we had stated that soft gamma-ray repeaters have been argued
to be young magnetars and that their birthrate is about 10\% of that
of the ordinary pulsars.  With one magnetar for every ten ordinary
pulsars, the probability of finding a middle-aged magnetar is small
but not negligible.  Observationally, if we associate \rxj\ and \rxjw\
with magnetars, the local number density of magnetars would be
comparable to that of middle-aged ordinary pulsars.  This would
indicate that magnetars are brighter and/or longer-lived as soft
\Xray\ sources.  Heyl and Hernquist (\cite{heyl:98}) have studied the
thermal evolution in the presence of magnetic fields in detail.
Curiously, their cooling timescales appear insensitive to the field
strength when $kT_{\rm{}eff}\lesssim100$\,eV.  However, their models
neglect the influence of magnetic-field decay.  For these high field
strengths, decay is expected (Thompson \& Duncan \cite{thomd:96}; also
Goldreich \& Reisenegger \cite{goldr:92}).  We believe this field
decay is what sets magnetars apart from and could make them brighter
than ordinary neutron stars.  Therefore, we would argue that \rxj\
could well be a magnetar.

\section{The nature of star X\label{sec:disc}}

We now discuss the nature of star X. There are two possibilities.  The
first is that X is unrelated to \rxj.  If so, X has to be a distant
star, with spectral type no later than F2 ($B-R<0.72$; 3$\sigma$
limit).  Combined with the apparent B magnitude of 26.6, it could only
be a white dwarf (a main sequence star would be well outside even the
extended halo of our Galaxy).  A white dwarf would have
$M_V\lesssim13.2$ (Allen \cite{alle:73}), and be at a distance
$\gtrsim\!4\,$kpc, $\gtrsim\!0.6\,$kpc above the plane.  It would have
to be a disk white dwarf, since white dwarfs in the halo would be too
cool.  The scale height of disk white dwarfs is $275\pm50$\,pc, and
the local density $0.6\,10^{-3}{\rm\,pc^{-3}}$ (Boyle \cite{boyl:89}).
At the latitude of \rxj, $-7\fdg8$, we would thus expect about one
white dwarf per 4 square arcmin.  Hence, the probability of finding
one within the 80 square arcsec HRI error box is $\sim\!6\,10^{-3}$.
We conclude that it is possible, though not very likely, that X is a
distant white dwarf.  A better color determination could settle this
issue; if the object really has $B-R<0$, the
putative white dwarf would have to be much brighter and thus be well
outside the disk.



%
%
%
%
%
%
%

This brings us to our second hypothesis, that star X is the optical
counterpart of \rxj.  The observed B and R magnitudes correspond to
fluxes of $92\pm14$ and $53\pm17\,$nJy, respectively.  Here, we used
the effective wavelengths and absolute calibration given by Bessel
(\cite{bess:92}): $\lambda_{\rm{}eff}(B,R)=0.436$, $0.638\,\mu$m,
$F_\nu(B,R\!=\!0)=4.00$, $3.06\,$kJy.  We can compare these fluxes
with those expected from the ROSAT data, by extrapolating the
best-fit, $kT_{\rm{}eff}=80\,$eV, black-body spectrum (see
Sect.~\ref{sec:rxj}).  Using the Rayleigh-Jeans formula,
$f_{\nu}=2\pi{}kT_{\rm{}eff}(\nu/c)^2(R/d)^2$, and
$(R/d)^2\simeq{}f_X/\sigma T_{\rm eff}^{4}$ -- where $f_X$ is the
unabsorbed flux in the ROSAT band (Sect.~\ref{sec:rxj}) -- the
predicted B and R-band fluxes are 18 and 8\,nJy, respectively.  The
observed optical flux is a factor of five larger. A similar
excess is seen in \rxjw\ (Walter \&\ Matthews \cite{waltm:97}).

There are several possible causes for the apparent excess.  Perhaps
the simplest is that the surface does not have a uniform temperature;
a two-component model, one component with $kT_1=80\,$eV and the other
with $kT_2\ll80\,$eV but occupying $4T_1/T_2$ times more area, would
satisfactorily account for both the optical and the \Xray\
measurements.  An alternative is that the excess is only apparent; the
emergent \Xray\ spectrum depends on the composition in the
neutron-star atmosphere and the effective temperature may thus well be
different from that inferred by fitting a black-body spectrum (Pavlov
et al.\ \cite{pavl&a:96}). Finally, the optical excess could be due
to non-thermal emission, similar to what has been inferred for PSR
B0656+14 (Pavlov, Welty, \& C\'ordova \cite{pavlwc:97}; Kurt et al.\
\cite{kurt&a:97}; Shearer et al.\ \cite{shea&a:97}) and Geminga
(Bignami et al.\ \cite{bign&a:96}).


The measurement of $\dot{P}$ would clearly be the most elegant way to
discriminate between the two models for \rxj.  For the magnetar, one
would expect steady spin-down, at a few $10^{-13}{\rm\,s\,s^{-1}}$
(Sect.~\ref{sec:magnetar}).  For the accreting old neutron star model,
Wang (\cite{wang:97}) estimates a spin-down rate
$\lesssim\!10^{-16}{\rm\,s\,s^{-1}}$.  This ignores the torques
exerted by the accreting matter, however, which might lead to either
spin-up or spin-down, at a much larger rate (Lipunov \& Popov
\cite{lipup:95}).  Other future observations (improved position and
better spectrum, AXAF; accurate optical and UV photometry, HST; radio
emission, VLA) will be of great interest no matter what the object
turns out to be.

While revising this manuscript, we became aware of a preprint by Motch
\&\ Haberl (\cite{motch:98}) reporting optical observations of \rxj.
They propose two candidates, X1 and X2.  X1 is our star X. X2 is about
3 arcseconds NW of X1 (see  Figure~\ref{fig:images}) and is nominally
outside the HRI error circles based on our astrometry. It is easily
detected in R but fainter than X1 in B. Given the abundance of such
faint red objects it is our opinion that X2 is most likely a background
object.

\acknowledgments SRK's research is supported in part by NASA and NSF.
The observations reported here were obtained at the W. M. Keck
Observatory, which is operated by the California Association for
Research in Astronomy, a scientific partnership among California
Institute of Technology, the University of California and the National
Aeronautics and Space Administration.  It was made possible by the
generous financial support of the W. M. Keck Foundation.  The Munich
Image Data Analysis System is developed and maintained by the European
Southern Observatory.

\end{document}